\documentclass[%
 reprint,
superscriptaddress,
 amsmath,amssymb,
 aps,
]{revtex4-2}

\usepackage{graphicx}
\usepackage{dcolumn}
\usepackage{bm}
\usepackage{url}

\begin{document}

\preprint{}

\title{Superconducting and structural properties of the type-I superconductor PdTe$_2$ \\under high pressure}

\author{Yusaku Furue}%
\affiliation{Graduate School of Science and Technology, Niigata University, 8050, Ikarashi 2-no-cho, Nishi-ku, Niigata, 950-2181, Japan}%
\author{Tadachika Fujino}%
\affiliation{Faculty of Science, Niigata University, 8050, Ikarashi 2-no-cho, Nishi-ku, Niigata, 950-2181, Japan}%
\author{Marc V. Salis}%
\affiliation{Van der Waals-Zeeman Institute, University of Amsterdam, Science Park 904, 1098 XH Amsterdam, The Netherlands}%
\author{Huaqian Leng}%
\affiliation{Van der Waals-Zeeman Institute, University of Amsterdam, Science Park 904, 1098 XH Amsterdam, The Netherlands}%
\affiliation{School of Physics, University of Electronic Science and Technology of China, Chengdu, 610054, China}%
\author{Fumihiro Ishikawa}%
\affiliation{Faculty of Science, Niigata University, 8050, Ikarashi 2-no-cho, Nishi-ku, Niigata, 950-2181, Japan}%
\author{Takashi Naka}%
\affiliation{National Institute for Materials Science (NIMS), Sengen 1-2-1, Tsukuba, Ibaraki 305-0047, Japan}%
\author{Satoshi Nakano}%
\affiliation{National Institute for Materials Science (NIMS), Sengen 1-2-1, Tsukuba, Ibaraki 305-0047, Japan}%
\author{Yingkai Huang}%
\affiliation{Van der Waals-Zeeman Institute, University of Amsterdam, Science Park 904, 1098 XH Amsterdam, The Netherlands}%
\author{Anne de Visser}%
\affiliation{Van der Waals-Zeeman Institute, University of Amsterdam, Science Park 904, 1098 XH Amsterdam, The Netherlands}%
\author{Ayako Ohmura}%
 \email[E-mail: ]{ohmura@phys.sc.niigata-u.ac.jp}%
\affiliation{Faculty of Science, Niigata University, 8050, Ikarashi 2-no-cho, Nishi-ku, Niigata, 950-2181, Japan}%

\date{\today}

\begin{abstract}
The transition metal dichalcogenide PdTe$_2$ has attractive features based on its classification as a type-II Dirac semimetal and the occurrence of type-I superconductivity, providing a platform for discussion of a topological superconductor. Our recent work revealed that type-I superconductivity persists up to pressures of $\sim2.5$ GPa and the superconducting transition temperature $T_{\rm c}$ reaches a maximum at around 1 GPa, which is inconsistent with the theoretical prediction. To understand its non-monotonic variation and investigate superconductivity at higher pressures, we performed structural analysis by x-ray diffraction at room temperature below 8 GPa and electrical resistivity measurements at low temperatures from 1 to 8 GPa. With regard to the superconductivity beyond 1 GPa, the monotonic decrease in $T_{\rm c}$ is reproduced without any noticeable anomalies; $T_{\rm c}$ changes from 1.8 K at 1 GPa to 0.82 K at 5.5 GPa with  $dT_{\rm c}/dP\sim-0.22$ K/GPa. The crystal structure with spacegroup $P$\={3}$m$1 is stable in the pressure range we examined. On the other hand, the normalized pressure-strain analysis (finite strain analysis) indicates that the compressibility changes around 1 GPa, suggesting that a Lifshitz transition occurs. We here discuss the effect of pressure on the superconducting and structural properties based on the comparison of these experimental results.     
\end{abstract}


\maketitle


\section{\label{sec:level1}Introduction}     
The family of transition metal dichalcogenides attracts much attention because of their versatile electronic properties. Notably some members are classified as a type-II Dirac semimetal, which implies a tilted Dirac cone, resulting in the breaking of Lorentz invariance, with a topologically non-trivial surface state \cite{Soluyanov2015, Huang2016, Bahramy2018}. Of particular interest in the family is PdTe$_2$, which is well known as a superconductor at ambient pressure \cite{Guggenheim1961}. The coexistence of characteristic electronic states with a non-trivial nature, which was experimentally confirmed \cite{Bahramy2018, Fei2017, Noh2017, Clark2018, Das2018, Amit2018_2}, in combination with superconductivity, enhances its reputation as a potential topological superconductor, and as such provides a platform for discussion of the relationship between superconducting properties and Dirac fermions. 

The superconducting properties of PdTe$_2$ have been widely investigated at ambient pressure \cite{Clark2018, Das2018, Leng2017, Amit2018, Salis2018, Teknowijoyo2018, Kim2018, Le2019, Voerman2019, Sirohi2019, Leng2019, Salis2021}. PdTe$_2$ is a type-I superconductor with 
$T_{\rm c}=1.64$ K and its critical field $\mu_0H_{\rm c}(T)$ follows the standard quadratic temperature variation with $\mu_0H_{\rm c}(0)=13.6$ mT  \cite{Leng2017}. The bulk superconducting property is found to be of conventional nature with a full superconducting gap confirmed by heat capacity \cite {Amit2018, Salis2021} and penetration depth measurements \cite {Salis2018, Teknowijoyo2018}. Furthermore, it is considered that a saddle-point van Hove singularity (vHs) near the $M$-point in the Brillouin zone, which is located about 30 meV above the Fermi level \cite{Erik2019}, plays an important role in the conventional Bardeen-Cooper-Schrieffer (BCS) behavior \cite{Kim2018}. Meanwhile, as regards the surface states of PdTe$_2$, its superconducting properties are under discussion \cite{Clark2018, Leng2017, Das2018, Le2019, Sirohi2019, Voerman2019}. Leng {\it et al.} observed an unusual surface sheath superconductivity in ac-susceptibility measurements in magnetic fields \cite{Leng2017}. Its critical temperature is estimated to be $T_{\rm c}^{\rm S} \sim1.3$ K and persists up to $\mu_0H_{\rm c}^{\rm S}(0)=$ 34.9 mT, which is higher than the bulk critical field $\mu_0H_{\rm c}(0)$. This anomalous behavior cannot be explained by the standard Saint-James--de Gennes model for surface superconductivity \cite{SJ-dG1963}. Specifically, in the standard theory, provided that the Ginzburg-Landau parameter $\kappa$ (= the penetration depth $\lambda$/coherence length $\xi$) is smaller than 0.42 in a type-I superconductor, the critical field of surface superconductivity should be smaller than that of the bulk. In the case of PdTe$_2$, despite $\kappa \sim 0.09-0.28<0.42$ \cite{Leng2017, Salis2018}, $\mu_0H_{\rm c}^{\rm S}(0)$ is larger than $\mu_0H_{\rm c}(0)$. This suggests that the surface sheath could have an unusual topological nature. In contrast, spectroscopic measurements such as scanning tunneling microscopy/scanning tunneling spectroscopy (STM/STS) and point-contact spectroscopy, which are genuine surface probes, reported that the surface superconductivity is conventional and no in-gap states exist \cite {Clark2018, Das2018, Voerman2019, Le2019, Sirohi2019}; different views on superconductivity are, however, type-II behavior \cite{Clark2018} or the mixture of type-I and type-II \cite{Le2019, Sirohi2019}. 

\begin{figure}[h]
  \begin{center}
    \includegraphics[clip,width=5.0cm]{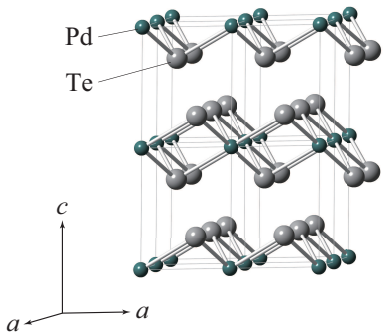}
    \caption{The crystal structure of PdTe$_2$ at ambient conditions with space group $P$\={3}$m$1 (No. 164). Green and grey spheres indicate the Pd and Te atoms, respectively.}
    \label{fig:1}
  \end{center}
\end{figure}

Since applying pressure can continuously tune structural and electronic states of materials, it is a most useful probe in research on superconductors and topological quantum materials. The effects of pressure on PdTe$_2$ have already been reported in several papers \cite {Soulard2005, Xiao2017, Leng2020, Leng2020SM, Yang2021}. PdTe$_2$ crystallizes in a layered structure with space group $P$\={3}$m$1 ($a= 4.037(2)$  \AA, $c= 5.132(2)$ \AA~\cite{Soulard2005} at ambient conditions) as shown in Fig. \ref{fig:1}. The Te-Pd-Te layers are weakly bonded by van der Waals interaction. Soulard {\it et al}. reported that the $P$\={3}$m$1 structure is maintained up to pressures of 27 GPa at room temperature \cite{Soulard2005}. First-principles calculations using the structural parameters of Ref. \cite{Soulard2005} predicted a monotonic depression of superconductivity up to pressures of 10 GPa, which is attributed to the decrease in density of states and the blueshift of the phonon frequency induced by pressure \cite{Xiao2017}. In our previous work, we found that the pressure variation of the transition temperature $T_{\rm c}(P)$ differs from that of the theoretical prediction; $T_{\rm c}$ reaches a maximum value of 1.91 K at 0.91 GPa and then decreases to 1.27 K at 2.49 GPa \cite{Leng2020}. To understand this result, we investigated the carrier density $n$ under pressure first. However, the variation of $n$ does not explain $T_{\rm c}(P)$ because $n$ shows only a slight increase  as a function of pressure \cite{Leng2020SM}. On the other hand, an enhancement of $T_{\rm c}$ is also reported in studies on Cu-intercalated Cu$_x$PdTe$_2$ ($T_{\rm c}^{\rm max}=2.6$ K for $x=0.06$) \cite{Hooda2018} and the Au-substituted series Au$_{1-x}$Pd$_x$Te$_2$ ($T_{\rm c}^{\rm max}=4.65$ K for $x=0.40$) \cite{Kudo2016}. We traced the relative change of $T_{\rm c}$ as a function of the relative volume change in Fig. 7 of Ref. \cite{Leng2020}. Though the value of $T_{\rm c}$ basically decreases with a smaller volume, the positive pressure variation of $T_{\rm c}$ up to 0.91 GPa is at odds with this trend and its origin remains unresolved.

Additionally, a pressure-induced topological transition is theoretically predicted at higher pressures by Xiao {\it et al.}; a type-I Dirac point appears at 4.7 GPa near the $\it \Gamma$ point and the type-II Dirac point near the $A$ point disappears at 6.1 GPa due to the pressure induced shifts of the electronic bands \cite{Xiao2017}. Very recently, Yang {\it et al.} reported that there is no anomalous behavior in the temperature dependence of the resistivity up to pressures of 8.2 GPa \cite{Yang2021}. However, the pressure variation of $T_{\rm c}$ is yet to be revealed in this pressure range since the data in their research were taken at $T > 2$ K. 

In this study, we focus on elucidating the origin of the non-monotonic variation of $T_{\rm c}$ and investigate superconductivity at higher pressures where the occurrence of a topological transition is expected. Furthermore, the detailed structural information below 2.5 GPa is also necessary to reveal the origin of the non-monotonic $T_{\rm c}(P)$, because the structural parameters of PdTe$_2$ below 2.5 GPa were not taken in Ref. \cite{Soulard2005}. For this purpose, the effects of pressure on the structural and superconducting properties of PdTe$_2$ up to pressures of 8 GPa were examined by synchrotron radiation x-ray diffraction and electrical resistivity measurements, respectively.

\section{Experiments}

The PdTe$_2$ crystals used in the present study were taken from the single-crystalline boule prepared under the modified Bridgman technique \cite{Lyons1976}. Scanning electron microscopy with energy dispersive x-ray (EDX) spectroscopy shows the proper 1:2 stoichiometry within the experimental resolution of 0.5\% (see Supplemental Material (SM) in Ref. \cite{Leng2017}).  The superconducting properties at ambient pressure were characterized in Ref. \cite{Leng2017}. 

A diamond anvil cell (DAC) was used for x-ray diffraction (XRD) under high pressure. A culet anvil of 0.75 mm in diameter was selected to fine tune the experimental pressure in the range below 8 GPa. A powder sample of PdTe$_2$ was prepared by grinding fragments cut out of a single crystal. It was placed in a sample space, 255  $\mu$m in diameter and 96 $\mu$m thick, made by drilling a small hole in a SUS301 gasket. We performed the XRD study twice utilizing different pressure tranmitting mediums. For hydrostatic compression, a 4:1 mixture of Methanol (MeOH) - Ethanol(EtOH) was loaded into the sample room together with the power sample in the first run, whereas fluid helium (He) compressed to $\sim180$ MPa was used in the second run \cite{Takemura2001}. XRD with synchrotron radiation was carried out at the beamline AR-NE1A of Photon Factory in the High-Energy Accelerator Research Organization (KEK) in Tsukuba, Japan. An incident beam was tuned to the energy of $\sim29.7$ keV ($\lambda \sim 0.417$ \AA) and beamsize of $75\times75~\mu$m$^2$. An imaging plate was used as a detector (Rigaku R-AXIS IV). All of the diffraction patterns under high pressures were measured at room temperature. Values of the applied pressures were determined using the ruby fluorescence method. The Rietveld analysis to obtain structural parameters was performed with the crystallographic program JANA2006 \cite{JANA2006}. A typical analytical result that gives a good fit is shown in Fig. S1 in the SM file.

Electrical resistivity measurements under high pressures and low temperatures were performed utilizing a Cu-Be modified Bridgman anvil cell  \cite{Nakanishi2002, Ishikawa2007, Nakanishi2007}. The PdTe$_2$ sample was put into a Teflon capsule together with a mixture of Fluorinert (FC70:FC77 = 1:1), which is a pressure-transmitting medium used for quasi-hydrostatic compression. The pressure value generated in the capsule against the load on the high-pressure cell was calibrated in advance by the critical pressures of the structural phase transitions in elemental bismuth (purity 5N). The PdTe$_2$ samples were cut out of a bigger single crystal; the typical size is $\sim0.3(W)\times0.7(L)\times0.1(H)~$mm$^3$. We measured the temperature variation of the resistivity, $\rho(T)$, of three samples under high pressure. These samples were compressed at room temperature and then the $\rho(T)$-curves at each pressure were measured under the following temperature conditions: sample 1 (the first run) at temperatures of $T\geq1.9$ K (in a physical property measurement system, PPMS, Quantum Design), sample 2 (the second run) at $T\geq0.8$ K (using a $^3$He-circulation--Joule-Thomson type Gifford-McMahon cryogenic refrigerator, Iwatani Industrial Gases Corporation), and sample 3 (the third run) at $T\geq0.3$ K (in a $^3$He refrigerator, Oxford Instruments Heliox VL).

\section{Results}
\subsection{X-ray diffraction}
\begin{figure}[t]
  \begin{center}
    \includegraphics[clip,width=7.0cm]{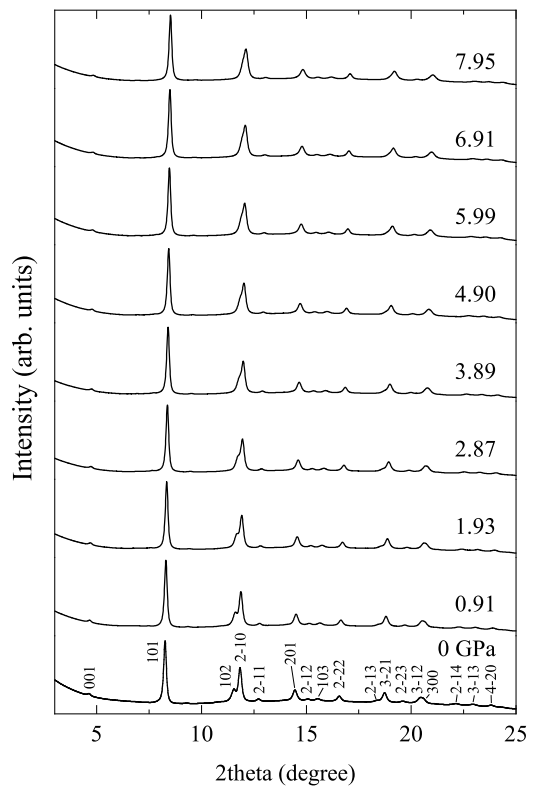}
    \caption{Diffraction patterns of PdTe$_2$ measured in the first run with increasing pressure at room temperature.}
    \label{fig:2}
  \end{center}
\end{figure}

Figure~\ref{fig:2} shows the pressure variation of the XRD patterns of PdTe$_2$ with increasing pressure taken in the first run. All reflections are indexed with $P$\={3}$m$1 symmetry. The diffraction pattern shifts to a higher scattering angle due to shrinkage of the crystal lattice and its profile is basically kept up to the highest pressure, indicating that no structural phase transition occurs in this range as reported in Ref. \cite{Soulard2005}. The 102 reflection merges into the 2\={1}0 one, suggesting that the $c$-axis of the $P$\={3}$m$1 structure is more compressible than the $a$-axis.

\begin{figure}[htbp]
  \begin{center}
    \includegraphics[clip,width=7.0cm]{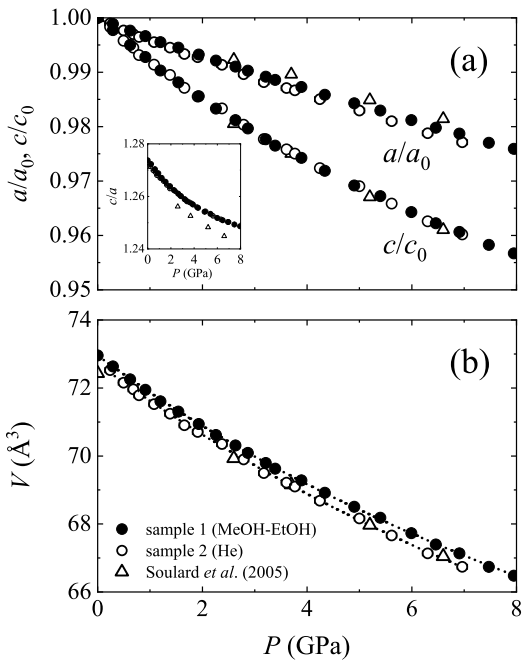}
    \caption{Pressure variation of the normalized lattice constants (a) and the unit-cell volume (b). Inset in Fig. \ref{fig:3}(a) shows the ratio $c/a$ as a function of pressure. Values shown as closed and open circles were obtained in the first and second run, respectively. Open triangles show the experimental values reported by Soulard {\it et al.} \cite{Soulard2005}. Broken lines in Fig. \ref{fig:3}(b) are the results of a fit to the BM-EOS (see text).}
    \label{fig:3}
  \end{center}
\end{figure}

Figure~\ref{fig:3}(a) shows the normalized lattice constants $a/a_0$ and $c/c_0$ as a function of pressure: $a_0= 4.0441(1)$ \AA~and $c_0= 5.1511(4)$ \AA~at 0 GPa measured in the first run. The experimental data obtained in this study are plotted together with those of Ref. \cite{Soulard2005} and their variations are consistent with each other. Both $a$ and $c$ monotonically decrease with pressure. The ratio of lattice constants $c/a$ shown in the inset of Fig.~\ref{fig:3}(a) also varies  without an extremum.  It suggests that the larger shrinkage of the crystalline lattice along the $c$-axis continues up to pressures of 8 GPa in PdTe$_2$. The compression curve can be fitted by the Birch-Murnaghan's equation of state (BM-EOS) as shown in Fig.~\ref{fig:3}(b) \cite{BMEOS1947}:
\begin{gather}
P=\frac{3}{2}B_0\left \{ \left( \frac {V_0}{V}\right)^{\frac{7}{3}}- \left (\frac {V_0}{V} \right)^{\frac{5}{3}} \right \} \cdot \nonumber\\ 
\left [ 1+\frac {3}{4}\left (B_0^{'}-4 \right ) \left \{  \left( \frac {V_0}{V} \right)^ {\frac{2}{3}}-1 \right \} \right], \nonumber
\end{gather}
where $V_0$ and $V$ are volumes at ambient and high pressures, $P$ is in units of GPa, $B_0$ is the bulk modulus, and $B^{'}_0$ its pressure derivative. The fits for the first and second runs give parameters $B_0=62.9(9)$ and 59(2) GPa and $B_0^{'}= 6.7(2)$ and 7.0(2), respectively, for PdTe$_2$ with the $P$\={3}$m$1 structure. The pressure error in the second run is a little larger than that of the first run, which is reflected in the error of $B_0$. Compared to Ref. \cite{Soulard2005}, the value of the bulk modulus in this study is smaller than theirs, $B_0=101.5$ GPa. This is attributed to the difference whether or not the volume at ambient pressure is fixed as a fitting parameter and the pressure range used for the fitting of the equation of state.

\begin{figure}[htbp]
  \begin{center}
    \includegraphics[clip,width=8.5cm]{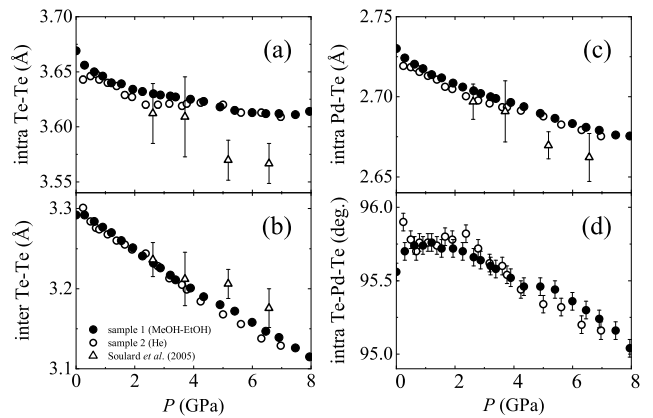}
    \caption{ Pressure dependences of the atomic spacings and bond angle in PdTe$_2$: (a) intralayer (intra) Te-Te, (b) interlayer (inter) Te-Te, (c) intra Pd-Te spacings, (d) intra Te-Pd-Te angle. Closed and open circles indicate values obtained in the first and second run, respectively. Open triangles show experimental values reported by Soulard {\it et al.} \cite{Soulard2005}.}
    \label{fig:4}
  \end{center}
\end{figure}

Figure~\ref{fig:4} shows the atomic spacings and bond angle of PdTe$_2$ with increasing pressure: (a) intralayer (intra) Te-Te, (b) interlayer (inter) Te-Te, (c) intra Pd-Te spacings, and (d) intra Te-Pd-Te angle. In PdTe$_2$, the atomic positions in the unit cell  are 1$a$ (0, 0, 0) for Pd and  2$c$ (1/3, 2/3, $z_{\rm Te}$) for Te, respectively; only $z_{\rm Te}$ can be optimized and its value at 0 GPa is $z_{\rm Te}=0.2747(5)$. The inter Te-Te and intra Pd-Te spacings shown in Fig. \ref{fig:4}(b) and \ref{fig:4}(c) indicate a monotonic decrease with pressure. Since the interlayer spacings in PdTe$_2$ are due to weak bonds by the van der Waals force,  the inter Te-Te spacing is about three times more compressible than other intralayer spacings. Meanwhile, the pressure variations of the intra Te-Te spacing and intra Te-Pd-Te angle show small non-monotonic changes. In particular, the angle hardly changes below $\sim2$ GPa and decreases beyond this pressure, though the values around 0.3 GPa in the first and second runs deviate slightly from the overall trend. From Table S1 in SM, it can be concluded that the value of $z_{\rm Te}$ approximately remains constant within the error up to $\sim1.2$ GPa and subsequently starts to increase with pressure. The intra Te-Pd-Te angle tends to increase up to $\sim1.2$ GPa, since the unit cell is more compressible along the $c$-axis. The extracted pressure variations of the atomic spacings are in agreement with literature data basically \cite{Soulard2005}. The difference in values above 4 GPa is probably caused by the degree of hydrostaticity, which is due to the solidification of the nitrogen used as a pressure-transmitting medium in Ref. \cite{Soulard2005}.

\subsection{Electrical resistivity}

\begin{figure}[htbp]
  \begin{center}
    \includegraphics[clip,width=8.0cm]{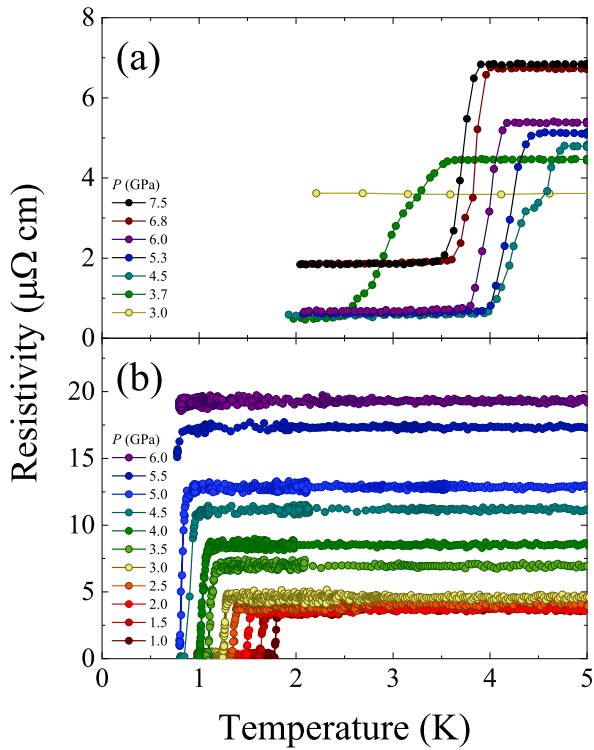}
    \caption{Temperature dependence of the electrical resistivity of PdTe$_2$ obtained in the first run (a) and second run (b) with increasing pressure up to 7.5 and 6.0 GPa, respectively. The measurement temperature ranges are (a) $T \geq$ 2K and (b)  $T \geq$ 0.8 K, respectively.}
    \label{fig:5}
  \end{center}
\end{figure}

We measured the pressure variation of the electrical resistivity on three samples. In the first run, shown in Fig.~\ref{fig:5}(a), a drop in the resistance corresponding to a  superconducting transition was observed above 3.7 GPa, though zero resistance was not attained. At this pressure, the transition is fairly broad and has three steps at 3.5, 3.0, and 2.6 K, suggesting an inhomogeneous superconducting state. On further compression, the superconducting transition shifts to higher temperatures at 4.5 GPa and is subsequently depressed and changes to a single-step transition up to pressures of 7.5 GPa. As mentioned in the Introduction, the appearance of the type-I Dirac point is theoretically predicted at 4.7 GPa. We, therefore, performed two more pressure runs to investigate the reproducibility of this higher-$T_{\rm c}$ phase. Figure~\ref{fig:5}(b) shows the pressure variation of the electrical resistivity obtained in the second run. A sharp transition is observed at $T_{\rm c}=1.8$ K at 1 GPa, which is the lowest experimental pressure in this run. The superconductivity is depressed with pressure and $T_{\rm c}$ decreases to 0.82 K at 5.5 GPa. Above $T_{\rm c}$ the normal state resistance is close to temperature independent and the anomalous behavior in the first run is not reproduced. 

\begin{figure}[htbp]
  \begin{center}
    \includegraphics[clip,width=7.5cm]{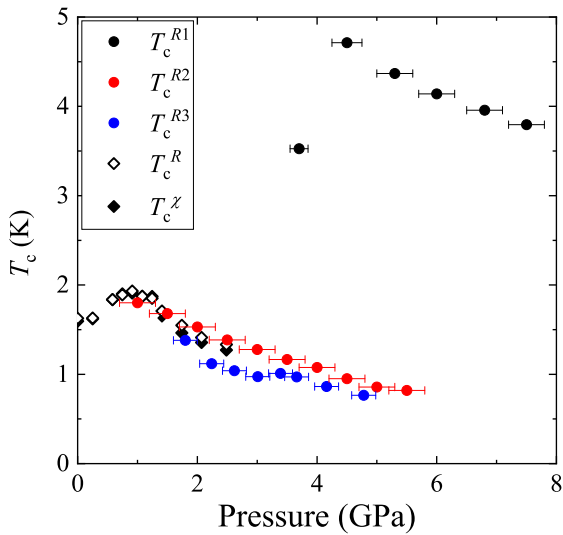}
    \caption{Pressure variation of the superconducting transition temperature of  PdTe$_2$. Black, red, and blue-closed circles indicate the onset temperatures obtained in the first ($T_{\rm c}^{R1}$), second ($T_{\rm c}^{R2}$), and third ($T_{\rm c}^{R3}$) run, respectively. Open and closed diamonds indicate $T_{\rm c}$ obtained in our previous resistivity ($T_{\rm c}^{R}$) and ac-susceptibility ($T_{\rm c}^{\chi}$) measurements \cite{Leng2020}.}
    \label{fig:6}
  \end{center}
\end{figure}

The pressure variation in the third run reproduces the one in the second run. Figure~\ref{fig:6} shows the onset temperatures of the superconducting transitions obtained in the three runs (and samples), $T_{\rm c}^{R1}$, $T_{\rm c}^{R2}$, and $T_{\rm c}^{R3}$, plotted together with $T_{\rm c}^{R}$ and $T_{\rm c}^{\chi}$ from our previous work \cite{Leng2020}. $T_{\rm c}^{R2}$ monotonically decreases with pressure derivative $dT_{\rm c}^{R2}/dP \sim-0.22$ K/GPa. $T_{\rm c}^{R3}$ also follows this variation under pressure, though between 2 and 3 GPa the data points lie a little lower. Furthermore, these values and pressure variation are consistent with those of $T_{\rm c}^{R}$ and $T_{\rm c}^{\chi}$. Therefore, it can be concluded that $T_{\rm c}^{R2}$ and $T_{\rm c}^{R3}$ must be attributed to the bulk superconducting phase of PdTe$_2$ and that the depression of this superconducting phase above 1 GPa is an intrinsic property. On the other hand, the superconducting phase with $T_{\rm c}^{R1}$ was observed only in the first run. Since the pressure variation of $T_{\rm c}^{R1}$ is clearly distinct, we infer that the superconductivity related to $T_{\rm c}^{R1}$ does not originate from PdTe$_2$, as will be discussed later.

\begin{figure}[t]
  \begin{center}
    \includegraphics[clip,width=8.5cm]{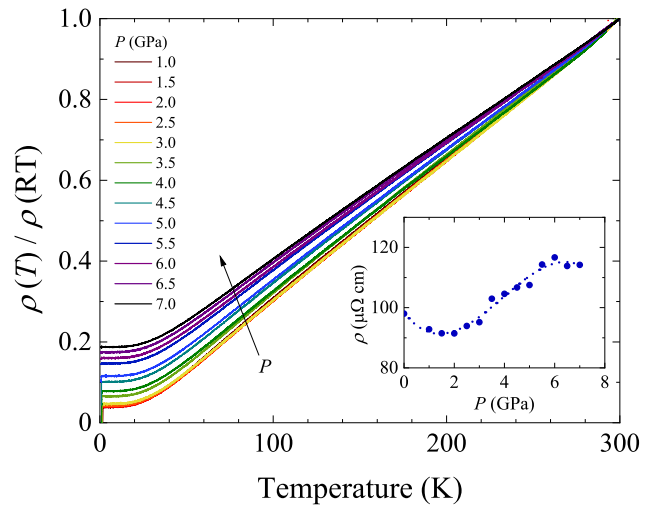}
    \caption{Pressure dependence of the electrical resistivity normalized to its value at room temperature $\rho$(RT) taken in the first run. Inset shows $\rho$(RT) as a function of pressure.}
    \label{fig:7}
  \end{center}
\end{figure}

Figure~\ref{fig:7} shows the electrical resistivity versus temperature normalized to its value at room tempeature, $\rho(T)/\rho(\rm{RT})$, obtained in the second run. The temperature dependence of the resistivity show metallic behavior, and no anomalies are observed up to 7.0 GPa. This behavior in  $\rho(T)/\rho(\rm{RT})$ is consistent with the one reported by Yang {\it et al.} \cite{Yang2021}.  The resistivity at room temperature is 98 $\rm{\mu \Omega}$ cm at ambient pressure and shows a minimum value around 2 GPa on compression as shown in the inset of Fig.~\ref{fig:7}. Furthermore, we estimated the Debye temperature ${\it \Theta}_D$ utilizing two sets of $\rho(T)$-data, which were obtained on sample 2 and another one (sample 4, $0.4 \leq P$ GPa $\leq 2.4$). Each $\rho(T)$-curve can be fitted to the Bloch-Gr\"{u}neisen formula based on electron-phonon scattering. ${\it \Theta}_D$ keeps values of $\sim184.8(6)$ K up to 1 GPa and then gradually increases with $\sim15$ K/GPa up to $\sim3$ GPa. A typical fitting result at 0.4 GPa on sample 4 and the pressure variation of ${\it \Theta}_D$ are shown in Fig. S2 and S3 in SM, respectively.

\section{Discussion}

First, we discuss the superconducting phase with $T_{\rm c}^{R1}$ that appeared above a pressure of 3.7 GPa in the first run. It is tempting to relate this higher-$T_{\rm c}$ phase to the occurrence of the pressure-induced topological transition from a type II to type I Dirac semimetal that was predicted by Xiao {\it et al.} \cite{Xiao2017}. At such a transition, the tilt parameter $k$ of the Dirac cone reaches a critical value of 1 and superconductivity is enhanced \cite{Shapiro2018}. However, as mentioned before, this higher-$T_{\rm c}$ phase has not been reproduced in the two subsequent runs. Furthermore, the pressure variation of the electrical resistivity at room temperature is also different for sample 1 and 2 as shown in Fig. S4 in SM; the resistivity of sample 1 monotonically decreases with pressure while that of sample 2 increases above 2 GPa as shown in the inset of Fig. \ref{fig:7}. These results suggest that the higher-$T_{\rm c}$ superconductivity is not caused by PdTe$_2$ but by another component. It is, therefore, concluded that the monotonic depression of the original superconductivity shown as $T_{\rm c}^{R2}$ and $T_{\rm c}^{R3}$ is the intrinsic behavior from 1 to 6 GPa in PdTe$_2$ and no indications related to the topological transition are observed in the resistivity measurements. 

Here we argue that $T_{\rm c}^{R1}$ can likely be attributed to the pressure-induced superconductivity of tellurium (Te). According to a past report \cite{Berman1973}, superconductivity in pure Te emerges above $P\sim4$ GPa and its pressure variation is very similar to that of $T_{\rm c}^{R1}$; $T_{\rm c}$ of pure Te increases from 3.4 K at $\sim4$ GPa to 4.3 K at $\sim6$ GPa and then gradually decreases to 2.8 K at $\sim15$ GPa \cite{Berman1973}. Possibly some trace amounts of elemental Te cause filamentary superconductivity under pressure. We remark, the Te content was below the threshold for the detection by EDX spectroscopy. The multi-step superconducting transition is compatible with this idea. The x-ray diffraction results under pressure indicate the amount of Te impurity phase must be very small, since no reflections related to Te have been observed.

Next, the non-monotonic variation of the superconducting transition temperature below 2 K, where $T_{\rm c}$  passes through the maximum of 1.91 K at 0.91 GPa, is discussed. In general, applying pressure tends to depress superconductivity. On the other hand, an increase in $T_{\rm c}$  under pressure has been observed in several transition metal chalcogenides, e.g. $MX_2$ ($M$=Ni, Pd, Nb, $X$=S, Se, Te) and ZrTe$_3$ \cite {Suderow2005, Li2021, ElGhazali2019, ElGhazali2017, Gu2018}. The effect of pressure on superconductivity has often been discussed based on the following points \cite{Lorenz2005, Schilling2007}: 1) the sign of $dT_{\rm c}/dP$ depends on the magnitude of the Hopfield parameter $\eta$ for electronic properties and the Gr\"{u}neisen parameter $\gamma$ for lattice hardening and 2) a variation of the energy bands near the Fermi energy ($E_F$) can contribute to a change in the density of states and also cause a pressure-induced electronic transition, which can alter the superconducting properties. 

We here consider the first point qualitatively. The sign of $dT_{\rm c}/dP$ is expected to become positive when the logarithmic volume derivative of $\eta=N(E_f)\langle I^2 \rangle$ is larger in magnitude than two times $\gamma=-d \ln\langle \omega \rangle/d{\ln}V$ \cite{Schilling2007}, where $N(E_f)$ is the electronic density of states at the Fermi level, $\langle I^2\rangle$ is the average square electronic matrix element, and $\langle \omega \rangle$ is the average phonon frequency. The carrier density in PdTe$_2$ at 2 K undergoes a quasi-linear increase by $\sim20$\% up to a pressure of $\sim2.1$ GPa without any anomalous behavior around 0.9 GPa, where $T_{\rm c}(P)$ has a maximum \cite{Leng2020SM}. Meanwhile, as mentioned before, ${\it \Theta}_D$, proportional to $\langle \omega \rangle$, changes little up to $\sim1$ GPa which results in a very small value of $\gamma$. From these two points,  the required condition for $dT_{\rm c}/dP>0$ might be fulfilled below 1 GPa. Beyond this pressure, the rise in ${\it \Theta}_D$ makes $\gamma$ larger overwhelming the moderate increase in $N(E_f)$ and thereby results in the depression of superconductivity. Based on this idea, the value of $d{\rm ln}\eta/d{\rm ln}V$ below 2.5 GPa can be estimated to be $\sim -2.9$ for $\gamma \sim 0$ at $P<1$ GPa and  $\sim -2.8$ for $\gamma \sim 3.3$ at $P>1$ GPa, respectively (see Sec. 4 in SM). The variation of $d\ln\eta/d\ln V$ is very small and consistent with those of carrier density.   

Next, we will examine the second point for the enhancement of superconductivity. Since PdTe$_2$ shows no structural phase transition up to pressures of 8 GPa, we explore possibilities of an electronic transition. For this purpose, a normalized pressure $H_v$ versus an Eulerian strain $f_E$ was calculated by employing the experimental lattice constants in the following formula,
\begin{gather}
H_v =\frac{P}{3f_E (1+2f_E)^\frac{5}{2}}, ~~~f_E=\frac{1}{2}\left [ \left( \frac{V_0}{V}\right)^\frac{2}{3} -1 \right ] \nonumber
\end{gather}
where $V_0$ and $V$ are volumes at ambient and high pressure, respectively, $P$ is in units of GPa, and
\begin{gather}
H_v=B_0 +\frac{3}{2}B_0 (B_0^{'}-4) f_E, \nonumber
\end{gather}
where $B_0$ is the bulk modulus, and $B^{'}_0$ its pressure derivative. As noted above, $H_v$ and $f_E$ have a linear relationship via $B_0$ and $B^{'}_0$. The intercept of the linear fit to $H_v(f_E)$ and the axis at $f_E=0$ gives an estimate of the bulk modulus in each region. Additionally, a change in the slope of $H_v(f_E)$ indicates a modification of the topology of the Fermi surface, suggesting that a Lifshitz transition occurs \cite{Lifshitz1960, Godwal2010, Polian2011}. 

\begin{figure}[b]
  \begin{center}
\includegraphics[clip,width=8.0cm]{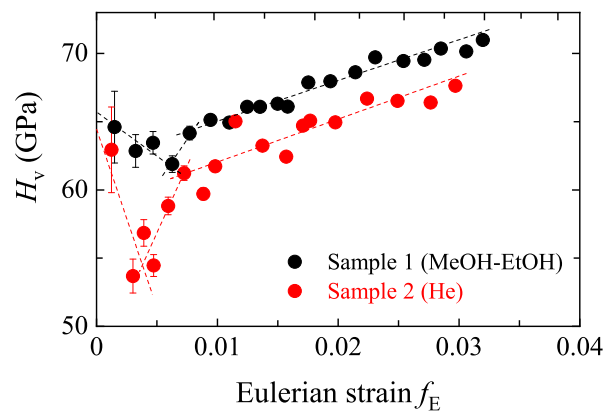}
\caption{Plot of the normalized pressure $H_v$ versus the Eulerian strain $f_E$ at ambient temperature. Black and red colors indicate data calculated from the first (sample 1) and second (sample 2) runs measured utilizing the mixture of MeOH-EtOH and He mediums, respectively. Broken lines show linear fits at each $f_E$ range.}
\label{fig:8}
  \end{center}
\end{figure}

Figure~\ref{fig:8} shows $H_v(f_E)$ obtained in two runs utilizing different pressure-transmitting mediums. The variation of $H_v(f_E)$ can be approximated by three straight lines with a different slope. This behavior is particularly noticeable in sample 2. A change in the slope occurs at $f_E \sim$ 0.006 ($P \sim$ 1.2 GPa) and  0.008 (1.6 GPa) for sample 1 and at $f_E \sim$ 0.004 ($P \sim$ 0.9 GPa) and 0.007 (1.3 GPa) for sample 2. Especially, the first variation around 1 GPa is very clear with a change from a negative slope to a positive one, suggesting that a Lifshitz transition occurs in PdTe$_2$. Intriguingly, the negative slope at the lowest $f_E$ range means that this range is stiffer than the higher $f_E$ ones: for example, $B_0=65.1(8)$ GPa at $f_E<0.006$, 54(2) GPa at $0.006<f_E<0.008$, and 62.3(3) GPa at $0.008<f_E$  in sample 1. Though this behavior is different from the general tendency under high pressure, we confirmed its reproducibility utilizing different mediums and, therefore, we expect it is to be intrinsic behavior.

Assuming that the topology of the Fermi surface changes around 1 GPa, we first consider the contribution of a saddle-point vHs, which sits $\sim30$ meV above $E_F$ near the $M$-point \cite{Erik2019}, to the transition. This is because a vHs is regarded as a key feature for the emergence of superconductivity in PdTe$_2$ \cite {Kim2018}. According to Kim {\it et al.}, the vHs band strongly correlates with the Te-shear $O_{1,2}$ phonon mode (in-plane vibrational mode), which tunes the intra Pd-Te spacing. They theoretically estimated the energy shift ($dE$) versus Te-displacement from the equilibrium position ($dx$) and the down shift of the vHs band to $E_F$ by shrinkage of the intra Pd-Te spacing. Referring to their $dx$ vs. $dE$ calculation as a rough guide, $dx$ of $\sim0.1$ \AA~is necessary to make the vHs band shift close to $E_F$. The estimated $dx$ is significantly larger than the shrinkage amount in intra atomic spacings around 1 GPa. Indeed, they claimed that the modification of the Fermi surface near the $M$-point will start at a 15\%-volume contraction \cite {Kim2018}, which corresponds to $P\sim17$ GPa using $B_0$ of this study. Therefore, combined with the variation of the carrier density, it is unreasonable to interpret that the energy shift of the vHs band causes a Lifshitz transition around 1 GPa and thereby enhances $T_{\rm c}$.  

Fermi surface sheets of PdTe$_2$ are also positioned near the ${\it \Gamma}$ and $K$-points \cite{Kim2018}, and it is speculated that $H_v(f_E)$ is affected by them, particularly at the ${\it \Gamma}$-point where the phonon modes provide a large contribution to superconductivity \cite{Kim2018}. Since the carrier density shows no remarkable changes around 1 GPa, we considered the factor of positive $dT_{\rm c}/dP$ based on the pressure variation of ${\it \Theta}_D$, which can also be linked to the change in $H_v(f_E)$. In general, for the same structure of the same substance, the pressure variation of the vibrational frequency tends to be smaller as the bulk modulus is higher. Therefore, the pressure variation of ${\it \Theta}_D$ can be compatible with the variation of $B_0$ estimated from $H_v(f_E)$. From a structural point of view, perhaps the variation of the Te-Pd-Te angle, which shows a weak non-monotonic behavior, leads to the change in compressibility. With regards to the superconducting properties, the pressure variations of the critical fields for the bulk and surface superconductivity, as well as those of $T_{\rm c}$, have a maximum around $0.9-1.2$ GPa \cite{Leng2020SM}. It suggests that the first variation of $H_v(f_E)$ around 1 GPa is reflected in superconductivity. Meanwhile, the second variation of $H_v(f_E)$ around 1.5 GPa also has the possibility of a Lifshitz transition. In our previous ac-susceptibility measurement, the critical temperature of the surface superconductivity becomes higher than that of the bulk above 1.41 GPa \cite{Leng2020}. This might be related to the electronic variation around 1.5 GPa in this study. However, for a more reliable discussion, it is necessary to obtain detailed information on the electronic structure and phonon dispersions at lower pressures utilizing experimental structural parameters in future research.

\section{Summary and conclusion}
We performed synchrotron radiation x-ray diffraction and electrical resistivity measurements on PdTe$_2$ under high pressure to investigate the effect of pressure on superconductivity up to 8 GPa and the origin of the non-monotonic variation of $T_{\rm c}$ observed in our previous study \cite{Leng2020}. With increasing pressure, $T_{\rm c}$ decreases from 1.8 K at 1 GPa to 0.82 K at 5.5 GPa with  $dT_{\rm c}/dP\sim-0.22$ K/GPa and the pressure variation below 2.5 GPa is consistent with our previous results. Though a high-$T_{\rm c}$ phase beyond 4 K was observed above 3.7 GPa in one of the samples, we concluded that this transition is likely due to contamination with Te. As for the topological transition from a type-II Dirac semimetal to type-I between 4.7 and 6.1 GPa theoretically predicted \cite{Xiao2017}, no noticeable behavior relating to the transition is observed in the pressure variation of both $\rho(T)$ and superconductivity. 

The pressure variation of $T_{\rm c}$ depends on the competition between the electronic density of states and lattice stiffening under pressure. In this study, we found that ${\it \Theta}_D$ estimated from $\rho(T)$ exhibits insignificant changes up to pressures of $\sim1$ GPa and subsequently increases as a function of pressure. Considering this together with the moderate increase in carrier density at 2 K with pressure \cite{Leng2020SM}, we infer that the requirement for positive $dT_{\rm c}/dP$ is fulfilled below 1 GPa and, however, large stiffening depresses superconductivity above 1 GPa. On the other hand, the $P$\={3}$m$1 structure is maintained up to pressures of 8 GPa similar to reported in previous literature \cite{Soulard2005}. As a result of our strain analysis, we found a clear change in compressibility around 1 GPa, suggesting the possibility that the Fermi surface is modified by pressure. It can be considered that this affects the pressure variation of ${\it \Theta}_D$, and eventually that of superconductivity. For a more solid discussion, further and more detailed information on the electronic structure of PdTe$_2$ at lower pressures is required.

\begin{acknowledgments}
 This work was partially supported by a Grant-in-Aid for Scientific Research, KAKENHI (No. 20H01851), the research program on Topological Insulators funded by FOM (Dutch Foundation for Fundamental Research on Matter), and performed under Proposal No. 2019G582 of Photon Factory, KEK. It was further supported by the JSPS (Japan Society for the Promotion of Science) Program for Fostering Globally Talented Researchers, Grant No. JPMXS05R2900003. 
\end{acknowledgments}


\newpage
\renewcommand{\baselinestretch}{1.2}
\setcounter{figure}{0}
\renewcommand*{\thefigure}{S\arabic{figure}}
\renewcommand*{\thetable}{S\arabic{table}}

\noindent \textbf{\Large  Supplemental Material for ``Superconducting and structural properties of the type-I superconductor PdTe$_2$ under high pressure''}

\vspace{1.0cm}
\noindent \textbf{1. Rietveld analysis of x-ray diffraction patterns and the list of structural parameters of PdTe$_2$}\\

\begin{figure}[htbp]
  \begin{center}
\includegraphics[clip,width=8.5cm]{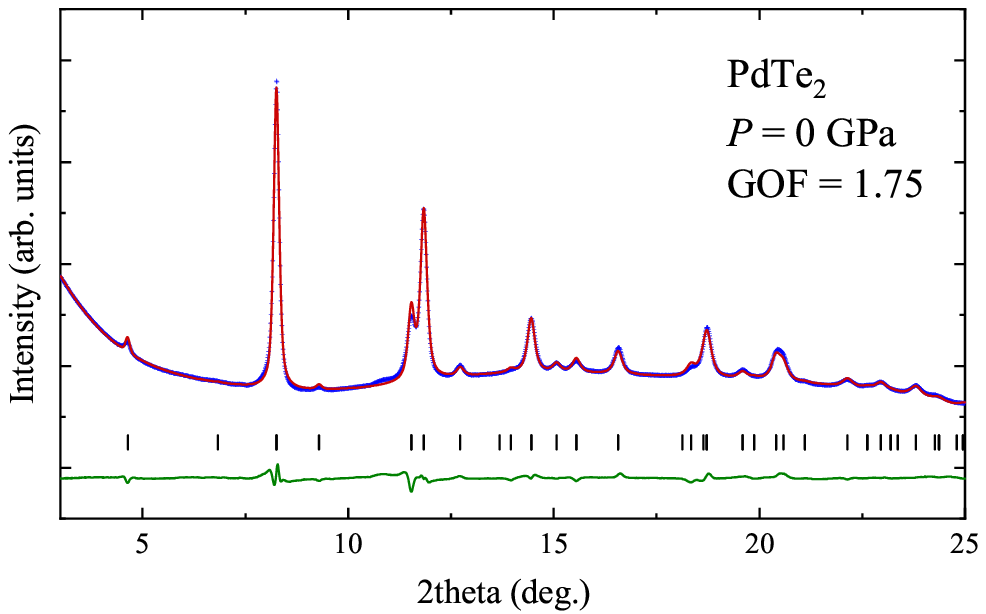}
\caption{X-ray diffraction pattern of PdTe$_2$ measured at $P=$ 0 GPa and the Rietveld refinement fitting results. Blue, red, and green lines indicate the experimental and calculated diffraction intensities, as well as the residual error. Small bars below the diffraction pattern indicate the diffraction angles in the $P\overline{3}m$1 structure. GOF means goodness of fit.}
\label{fig:s1}
  \end{center}
\end{figure}

The Rietveld analysis was performed using the crystallographic program JANA2006 in order to obtain the structural parameter \cite{Rietveld1967, JANA2006}. Figure~\ref{fig:s1} shows a typical result for one of the diffraction patterns measured at $P=$ 0 GPa. It demonstrates the quality of the fit. All diffraction patterns obtained in this study were analyzed with a similar accuracy. Table \ref{tab:1} shows the list of structural parameters obtained at each pressure. Values in this table were obtained in the first run with a 4:1 mixture of Methanol (MeOH)-Ethanol (EtOH) used as pressure transmitting medium. Since the MeOH-EtOH medium maintains a liquid condition up to 10 GPa, the hydrostaticity in the sample space can be kept at the maximum pressure of 8 GPa in this study. The lattice parameters $a$(\AA), $c$(\AA), and the unit cell volume $V$(\AA$^3$), monotonically decrease with pressure. The $z$-position of Te ($z_{\rm Te}$) remains constant within the error up to $\sim$ 1 GPa and then starts to increase as a function of pressure.

\vspace{0.5cm}

\begin{table}[htbp]
\caption{\label{tab:1} Structural parameters obtained at each pressure: lattice constants $a$(\AA) and $c$(\AA), volume $V$(\AA$^3$) of the unit cell, and the atomic $z$-position of  Te ($z_{\rm Te}$).  In the $P\overline{3}m$1 structure for PdTe$_2$ ($Z$=2), the atomic positions are (0, 0, 0) for Pd and (1/3, 2/3, $z_{\rm Te}$) for Te. As regards the atomic position, only the value of $z_{\rm Te}$ can be optimized by the Rietveld analysis.}
\begin{ruledtabular}
\begin{tabular}{ccccc}
\mbox{Pressure}&\mbox{$a$}&\mbox{$c$}&\mbox{$V$}&\mbox{$z_{\rm Te}$}\\
\mbox{(GPa)}&\mbox{(\AA)}&\mbox{(\AA)}&\mbox{(\AA$^3$)}&\mbox{}\\
\hline\\
0  & 4.0441(1) & 5.1511(4) & 72.957(5)  & 0.2747(5)   \\
0.29  &4.0396(1) & 5.1395(4) & 72.634(5)  & 0.2740(5)  \\
0.62 & 4.0346(1) & 5.1254(4)  & 72.255(5) & 0.2741(5)  \\
0.91  & 4.0305(1) & 5.1139(4)  & 71.947(5) & 0.2745(5)  \\
1.20  & 4.0260(1)  & 5.1012(4) & 71.607(5)  & 0.2746(5)  \\
1.54  & 4.0219(1) & 5.0899(4)  & 71.304(5) & 0.2752(5) \\
1.93  & 4.0169(1)  & 5.0768(4)   & 70.942(5) & 0.2756(5)   \\
2.26 & 4.0123(1)  & 5.0652(4)   & 70.617(5) & 0.2762(5) \\
2.63 & 4.0079(1)  & 5.0542(4)  & 70.311(5) & 0.2767(5)  \\
2.87 & 4.0048(1)  & 5.0463(4)   & 70.091(5) & 0.2771(5)  \\
3.21 & 4.0004(1)  & 5.0361(4)  & 69.796(5) & 0.2777(5) \\
3.39 & 3.9980(1)  & 5.0301(4)  & 69.628(5) & 0.2780(5)  \\
3.89 & 3.9927(1)  & 5.0184(4)  & 69.283(4) & 0.2787(5)  \\
4.34 & 3.9870(1)  & 5.0062(4)  & 68.917(4) & 0.2795(5)  \\
4.90 & 3.9806(1)  & 4.9922(4)  & 68.506(4) & 0.2799(5)  \\
5.40 & 3.9752(1)  & 4.9823(4)  & 68.185(4) & 0.2803(5)  \\
5.99 & 3.9680(1)  & 4.9672(4)  & 67.731(4) & 0.2813(5)  \\
6.46 & 3.9625(1)  & 4.9566(4)  & 67.398(4) & 0.2820(5)  \\
6.91 & 3.9580(1)  & 4.9484(3)  & 67.137(4) & 0.2826(5)  \\
7.47 & 3.9512(1)  & 4.9362(3)  & 66.738(4) & 0.2835(5)  \\
7.95 & 3.9468(1)  & 4.9281(3)   & 66.480(4) & 0.2845(5)  \\
\end{tabular}
\end{ruledtabular}
\end{table}

\vspace{0.2 cm}
\noindent \textbf{2. Estimation of the Debye temperature ${\it \Theta_D}$ utilizing the Bloch-Gr\"{u}neisen formula }\\

\begin{figure}[h]
  \begin{center}
\includegraphics[clip,width=8.0cm]{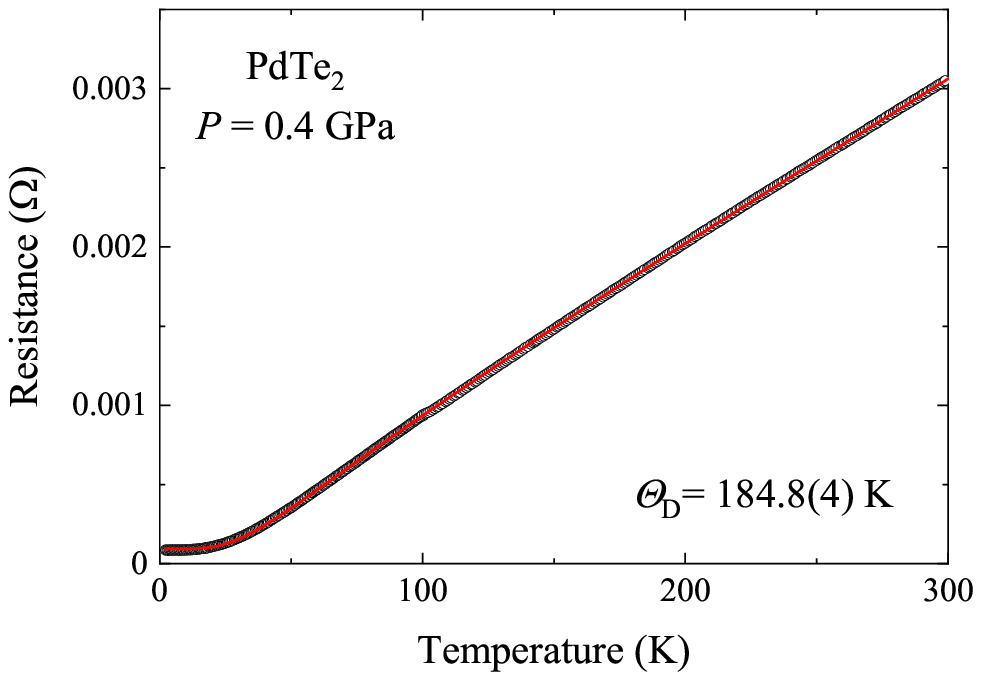}
\caption{The result of Bloch-Gr\"{u}neisen fitting on the resistance of sample 4 measured at 0.4 GPa. The dimension of sample 4 is $\sim1.4(W){\times}2.7 (L){\times}0.05 (H)$ mm$^3$ used in the piston-cylinder cell. The black symbols and red line indicate the experimental data and fitting result, respectively.}
\label{fig:s2}
  \end{center}
\end{figure}

\begin{figure}[h]
  \begin{center}
\includegraphics[clip,width=8.0cm]{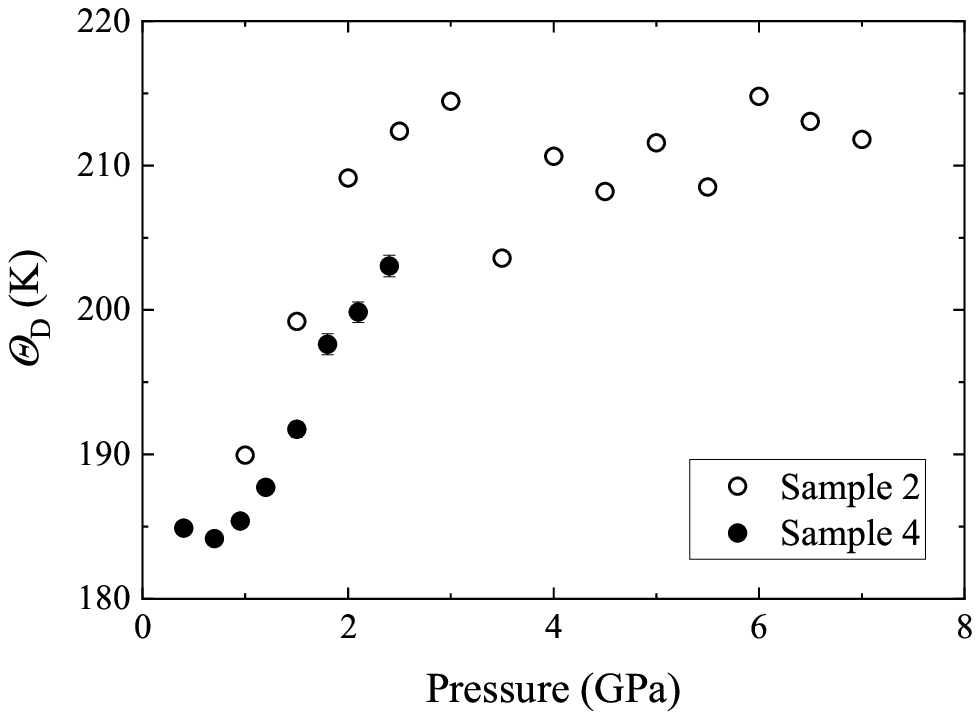}
\caption{Pressure variations of the Debye temperature  ${\it \Theta}_D$ estimated from the experimental resistivity data. Open and closed circles indicate values of ${\it \Theta}_D$ obtained on samples 2 and 4, respectively.}
\label{fig:s3}
  \end{center}
\end{figure}

We estimated the pressure variation of the Debye temperature ${\it \Theta_D}$ from the Bloch-Gr\"{u}neisen (BG) formula. The $\rho(T)$ data can be fitted to the BG formula based on an electron-phonon scattering with temperature exponent $n$=5:
{\small
\begin{eqnarray}
\rho(T) =\rho_0 + 4.225 \rho_{\it \Theta} \left (\frac{T}{\it \Theta_D}\right)^5 \int_0^{{\it \Theta_D}/T} \frac {x^5}{(e^x-1)(1-e^{-1})}dx, \nonumber
\end{eqnarray}
}
where $\rho_0$ is the residual resistivity, ${\it \Theta}_D$ is the Debye temperature, and $\rho_{\it \Theta}$ is the resistivity at ${\it \Theta}_D$. Besides for sample 2, which is introduced in the main manuscript, we fitted $\rho(T)$ of another sample (sample 4), which was measured using a piston-cylinder cell in the PPMS, to obtain values of ${\it \Theta}_D$ below 2.5 GPa in detail. Figure \ref{fig:s2} shows the fitting result at 0.4 GPa for sample 4. It results in ${\it \Theta}_D$ = 184.8(4) K. Other data of samples 2 and 4 were also fitted to the BG formula with the same fitting quality. Figure \ref{fig:s3} shows the pressure variation of ${\it \Theta_D}$ obtained for samples 2 (open symbols) and 4 (closed symbols). From the results of sample 4, the value of ${\it \Theta}_D$ seems  to remain almost constant at first and then rapidly increases beyond 1 GPa. The rapid increase of ${\it \Theta}_D$ between 1 and 3 GPa is also obtained for sample 2.

\vspace{1cm}

\vspace{0.5cm}
\noindent \textbf{3. Pressure variation of the electrical resistivity of PdTe$_2$ at room temperature}\\

\begin{figure}[t]
  \begin{center}
\includegraphics[clip,width=7.5cm]{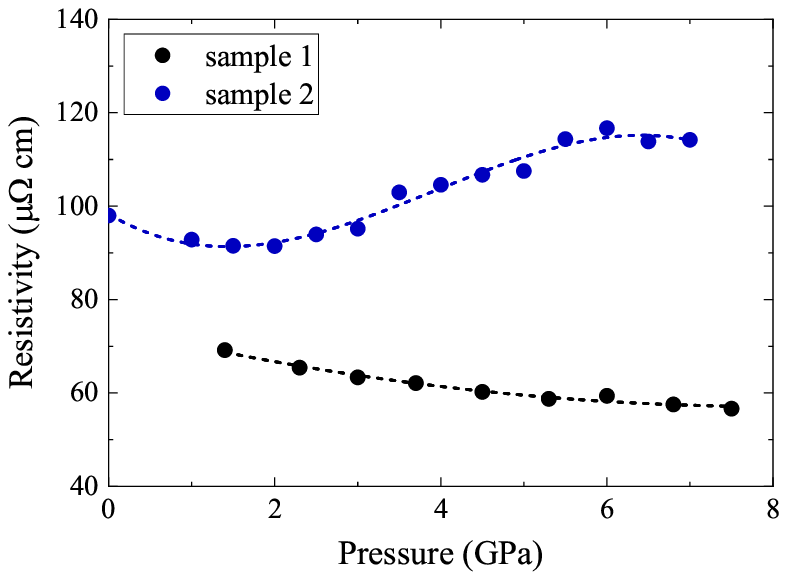}
\caption{Pressure variation of the electrical resistivities of samples 1 and 2 measured at room temperature. Black and blue circles indicate values obtained for samples 1 and 2, respectively. Broken lines are guides for the eye.}
\label{fig:s4}
  \end{center}
\end{figure}

Figure \ref{fig:s4} shows the value of the electrical resistivity at room temperature obtained for samples 1 and 2 as a function of pressure.  The observed pressure variations disagree with each other, that is to say, the resistivity of sample 1 monotonically decreases with pressure, while that of sample 2 changes the tendency from downward to upward between 1 and 2 GPa. As presented in the main manuscript, the appearance of superconductivity in samples 1 and 2 is also different as shown in Fig. 5 (a) and (b). We infer that sample 1 contains a small amount of elemental Te. This possibly also affects the room temperature resistance values.  

\vspace{0.5cm}

\vspace{0.5cm}
\noindent \textbf{4. Estimation of the logarithmic volume derivative of the Hopfield parameter $\eta$}\\

Referring to Ref. \cite{Schilling2007}, the logarithmic volume derivative of $T_{\rm c}$ derived from the Bardeen-Cooper-Schrieffer (BCS) expression is shown as follows:

\begin{eqnarray}
\frac{d\ln T_{\rm c}}{d{\ln}V} &=& -B_0 \frac {d\ln T_{\rm c}}{dP} =-\frac {B_0}{T_{\rm c}} \frac {dT_{\rm c}}{dP} \nonumber\\
&=& -\gamma +\left[ \ln \left(\frac{{\it \Theta}_D}{T_{\rm c}}\right)\right]\left[\frac{d\ln\eta}{d{\ln}V}+2\gamma \right],\nonumber
\end{eqnarray} 

\noindent where $T_{\rm c}$ is the superconducting transition temperature, $V$ is the volume, $B_0$ is the bulk modulus, $P$ is in units of GPa, $\eta~(=N(E_f)\langle I^2 \rangle)$ is the Hopfield parameter, $\gamma~(=-d \ln\langle \omega \rangle/d{\ln}V)$ is the Gr\"{u}neisen parameter, and ${\it \Theta}_D~(=\langle \omega \rangle/0.83)$ is the Debye temperature. The sign of $dT_{\rm c}/dP$ is expected to become positive when the logarithmic volume derivative of $\eta$, $d\ln\eta/d{\ln}V$, is larger in magnitude than $2\gamma$. 

In the pressure range up to 2.5 GPa, we estimated the value of $d\ln\eta/d{\ln}V$ below and above 1 GPa utilizing the experimental parameters, $T_{\rm c}$, $B_0$, and ${\it \Theta}_D$, obtained in this study and Ref. \cite{Leng2020}. $\gamma$ was estimated from the pressure variation of ${\it \Theta}_D$ obtained on sample 4 (see Fig. S3): $\gamma \sim0$ at $P < 1$ GPa and  $\sim 3.3$ at $P>1$ GPa. The calculated value for $\ln({\it \Theta}_D/T_{\rm c})$ was $\sim4.6$ at 0.4 GPa and $\sim5.0$ at 2.4 GPa. Combining these values with $B_0=62.9$ GPa and $d\ln T_{\rm c}/dP$, the value of $d{\rm ln}\eta/d{\rm ln}V$ below 2.5 GPa can be estimated to be $\sim -2.9$ at $P<1$ GPa and  $\sim -2.8$ at $P>1$ GPa, respectively.


\end{document}